# High temperature spin-freezing transition in Pyrochlore $Eu_2Ti_2O_7$: A new observation from ac-susceptibility

Arkadeb Pal[1], Abhishek Singh[1], Prajyoti Singh[1], A. K. Ghosh[2] and Sandip Chatterjee[1,*]

[1] Department of Physics, Indian Institute of Technology (Banaras Hindu University), Varanasi-221005, India

[2]Department of Physics, Banaras Hindu University, Varanasi-221005, India

## Abstract

The magnetization (both dc and ac) of pyrochlore $Eu_2Ti_2O_7$ have been investigated. Ac susceptibility ($\chi^{/}$&$\chi^{//}$) measurements reveal a new unusual spin freezing below 35K for pure $Eu_2Ti_2O_7$. Frequency dependence of these ac $\chi^{//}$peak positions indicate slow spin relaxation near this temperature and it follows the Arrhenius formulasuggesting a thermally activated relaxation process. The origin of this spin freezing has been attributed essentially to a single ion process which is associated to $Eu^{3+}$ spin relaxation. Non-magnetic dilution by $Y^{3+}$ ions also confirms the single ion freezing.

*Corresponding author e-mail id: schatterji.app@iitbhu.ac.in

## Introduction

Pyrochlore oxides being geometrically frustrated magnetic materials have attracted intense research attention due to its geometrical spin frustration driven novel ground states [1-3]. This plethora of interesting ground states is picked up by proper balance between exchange interaction, dipolar interaction and strong crystal field effect (CF). These special low temperature states include a fluid like states of spins, called spin liquid [4-6], spin glass,[7-9] spin ice [2,3,10-12], and order by disorder [13,14,15] etc. In pyrochlore $R_2Ti_2O_7$, the rare earth ions $R^{3+}$, occupying lattice points of corner shared tetrahedral play the main role in deciding the magnetic properties of the system. The R and Ti ions are individually occupying the lattice points of corner shared tetrahedra which collectively form inter-penetrating network of these tetrahedra, leading to frustration of their antiferromagnetic interactions. The unit cell of such structure contains 8 oxygen ions surrounding each $R^{3+}$ ion inside a trigonally distorted cube, two of these are situated in diametrically opposite to each other along the <111> axis ($D_{3D}$) and other six are lying on the equatorial plane of the cube [16]. The $R^{3+}$ ions are located at the vertices of two diametrically opposed tetrahedra of $R^{3+}$ ions i.e they are all having six nearest neighbours [16]. Strong crystal field interaction with the $D_{3D}$ symmetry causes the $R^{3+}$ ionic magnetic susceptibility to be different along the $D_{3D}$ axis and perpendicular to it, giving rise to Single ion anisotropy (SIA) [17]. In spin ice compounds ($Dy_2Ti_2O_7$ and $Ho_2Ti_2O_7$etc.), an unusual ground state is achieved by spins which is explained by Pauling's ice rule [10,18,19]. In spin ice materials, the f – electron spins of the rare earth ions $R^{3+}$ (R=Dy, Ho) are large and hence treated classically and the CF driven Single ion anisotropy (SIA) renders the spins to be Ising like along the <111> axes [18,19]. Under this condition, the system does not become ordered by minimizing the dipolar interactions alone, thus end up with a frozen, non-collinear and disordered state at very low temperature $T<T_{ice} \sim 4K$ [10,19]. But for $Dy_2Ti_2O_7$, an additional unexpected peak in ac susceptibility at ~ 16K is found, which is absent in spin ice $Ho_2Ti_2O_7$ suggesting a strange difference between these compounds [10,12]. This spin freezing at T=16K >$T_{ice}$ is attributed to the single spin freezing process [12,19-23]. The ice freezing (<4K) and this 16K single ion freezing are inter-linked by quantum tunnelling process (which is characterized by a very weak temperature dependence of spin relaxation times thus showing a plateau region below 12K) through the CF barrier and this is explained by creation and propagation of monopoles [19,24].

In this paper, we report temperature dependent magnetic study of the pyrochlore $Eu_2Ti_2O_7$. The special $4f^6$ configuration of $Eu^{3+}$ ion in $Eu_2Ti_2O_7$, plays the key role in determining its electronic and magnetic properties. In the unit cell of $Eu_2Ti_2O_7$,each $Eu^{3+}$ ion is surrounded by 8 oxygen atoms which form a trigonally distorted cube having a three-fold inversion $D_{3D}$ symmetry. The interactions between the $Eu^{3+}$ ions and surrounding 8 oxygen ions produce a crystal field with $D_{3D}$ symmetry [25]. The strong spin-orbit coupling in $Eu^{3+}$ ions results in forming its electronic pattern consisting of non-magnetic ground state $^7F_0$ and the first excited magnetic state $^7F_1$whichlies closely above it. This first magnetic term $^7F_1$ is followed by other excited magnetic levels $^7F_{2-6}$lying above $^7F_1$ successively. Again crystal field (CF) level splits into further levels e.g$^7F_1$ splits into a singlet and a doublet [25]. Thus, instead of having non-magnetic ground state, $Eu^{3+}$ shows appreciable magnetic susceptibility. In previous reports on $Eu_2Ti_2O_7$, CF calculations showed planar single ion anisotropy (SIA) parallel to local<111>axes [25]. As SIA along with exchange and dipolar interactions was seen to produce exotic magnetic properties, it was motivational for us to investigate detailed ac and dc magnetic study of this system. To our utter surprise, we observe a prominent spin freezing transition in ac susceptibility, in the form of a sudden drop in $\chi^{/}$ (real part) and the corresponding frequency dependent peaks in $\chi^{//}$ below 35K. Most interestingly, observation of spin freezing transition at such a higher temperature was not reported earlier in other pyrochlores, thus the finding becomes significantly important. Existence of such higher temperature freezing transition is extremely unusual in pure pyrochlores as they almost don't have any structural or chemical disorder (<1%) which is the origin of spin glass behaviour [12]. Hence the underlying freezing mechanism is associated to the geometrical frustration of rare earth magnetic spins and the anisotropy of the spins originated from CF of the system. Details discussions of the observed magnetic behaviours and its possible origin have been followed in this paper.

## *Experimental Details*

Polycrystalline samples of $Eu_2Ti_2O_7$ and $EuYTi_2O_7$ were synthesized using conventional solid state reaction method. High purity (>99.99%) $Eu_2O_3$, $Y_2O_3$ and $TiO_2$ were mixed in stoichiometric ratio and ground for 0.5 hr, then heated in air at $1000^0$ C for 24 hours. The resulting powder was reground and pressed into pellets and heated in air at $1250^0$ C for 48 hours and the process was repeated several times. X-ray diffraction measurement was performed using RigakuMiniflex II X-ray diffractometer. Figure. 1 shows the X-ray diffraction (XRD) data collected at room temperature (300K) along with its Rietveld refinement for pure $Eu_2Ti_2O_7$ sample. The XRD pattern was refined with space group Fd-3m. The position ofEu, Ti, O1 and O2 are 16d, 16c, 48f, 8c respectively. It suggests the samples are of good quality and confirms absence of chemically impure phase.We have performed ac and dc magnetic measurementsusing a Quantum Design magnetic property measurement system (MPMS) super conducting quantum interference devices (SQUID) magnetometer.

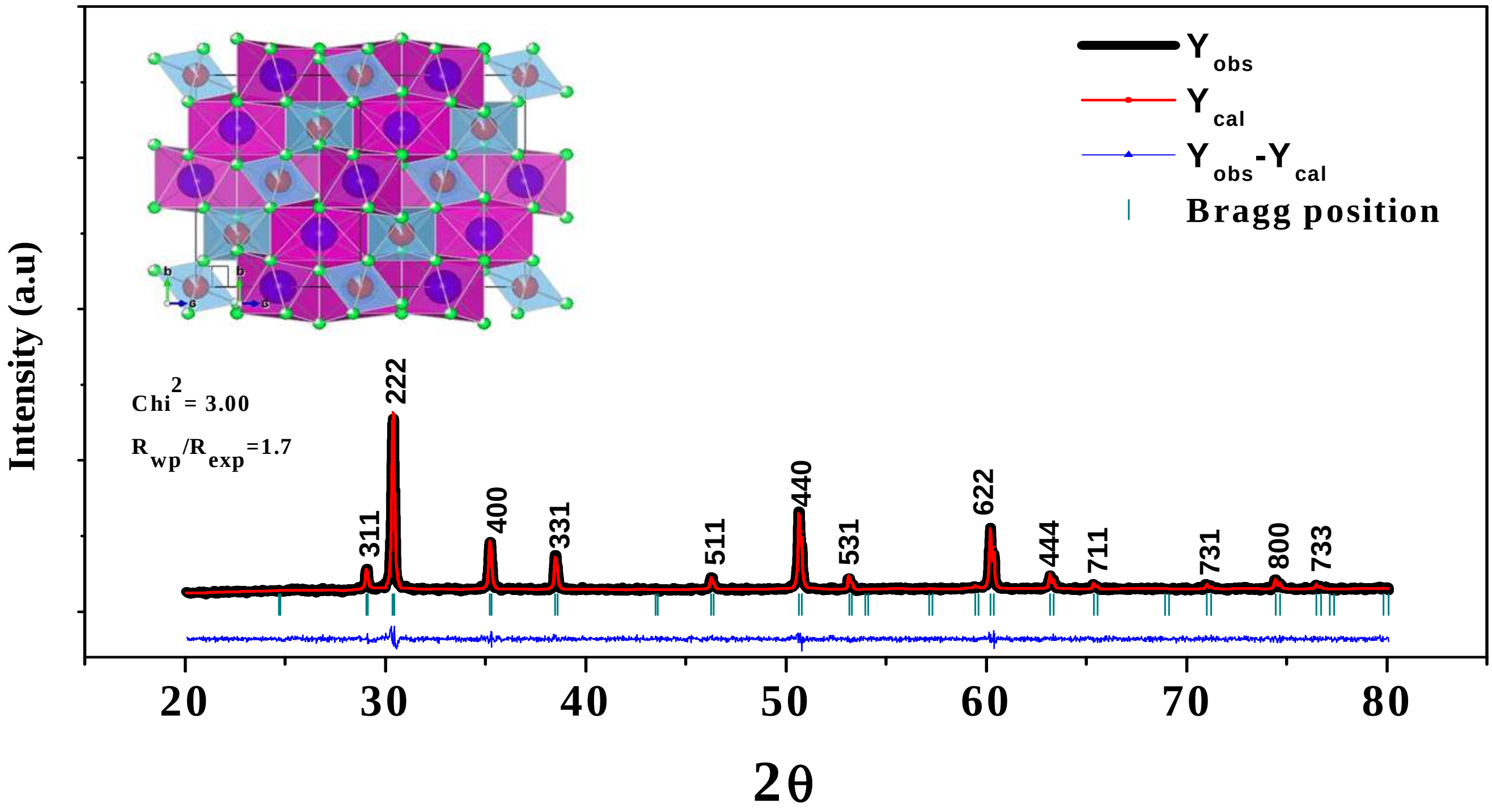

Fig.1

## Results and Discussions

### *Dc magnetization studies*

Dc magnetization measurement of $Eu_2Ti_2O_7$was performed as a function of temperature and field. Fig. 2 shows the temperature (T) variation of magnetization (M) following the zero field cooled (ZFC) and field cool (FC) protocols with applied magnetic field of 100 Oe. It shows magnetization (M) increases with decrease in temperature, however, there are few observations: As the temperature decreases from 300K M increases and reaches a maximum at ~90K. With further decrease of temperature, a slope change is observed and the system enters in a plateau region in which with temperature very weak variation in magnetization is observed. The plateau region extends down to ~20K and below of which again the slope of the curve changes and a sharp increase in magnetization is observed. The plateau region accounts for the strong crystal field effect (CF) [25]. The sharp increase in M below 10K may indicate some type of magnetic ordering. However, both the ZFC and FC magnetization curves have the same nature but a small difference is visiblebetween these two curves, suggesting existence of spin frustration in this system. However, the sharp increase in magnetization at low temperature cannot be attributed to the crystal field effect instead other magnetic interactions, e.g. exchange interactions, dipolar interactions etc., can be possible origin behind such behaviour. Therefore, to calculate the contributions of these different magnetic properties, high temperature series expansion of the susceptibility **$(\chi) = C\,[\frac{1}{T} + \frac{\Theta_{cw}}{T^2}]$** is considered [25,26]. We first plot the **$\chi T$** as a function of $\frac{1}{T}$ and then calculate the value of Curie Weiss temperature $\Theta_{cw}$, effective magnetic moment $\mu_{eff}$, exchange interaction energy $J_{nn}$ and dipolar interaction energy $D_{nn}$ from linear fit of the curve(inset (b) of figure 2). The fit has been done in the temperature range 2-5 K. Here, $\mu_{eff}$ is determined from $C= \frac{N\mu_{eff}^2}{3K}$; exchange interaction energy is obtained from $J_{nn} = \frac{3\Theta_{cw}}{zS(S+1)}$ , z=6 is the co-ordination number and dipolar interaction energy is determined from $D_{nn}= \frac{\mu_{eff}^2\mu_0}{4\pi r_{nn}^3}$ , here $r_{nn}$ is the distance between a $Eu^{3+}$ ion at (000) and its nearest neighbour at (*a*/4,*a*/4,0), ***a*** being the lattice constant of the unit cell [25,27]. The evaluated values of all the parameters ($J_{nn}$, $D_{nn}$, $\mu_{eff}$, $\Theta_{cw}$) obtained from aforementioned formulae have been summarized in Table-1. The data for pure $Eu_2Ti_2O_7$ shows nearest neighbour AFM exchange interaction $J_{nn}$ dominates over FM dipole-dipole interaction $D_{nn}$ and Curie-Weiss temperature is small but negative(-1.35K) which is consistent with the previous report [25]. Isothermal magnetization (M) as a function of magnetic field (H) at 2K (inset of fig. 2)shows the linear behaviour suggesting antiferromagnetic ordering of the system.

**Table-1**: **Showing the magnetic characteristic parameters evaluated from high temperature series expansion of susceptibility study for the temperature range 2K-5K.**

| Sample | Curie-Weiss temperature($\Theta_{cw}$) | Effective magnetic moment ($\mu_{eff}$) | Exchange interaction energy ($J_{nn}$) | Dipolar exchange interaction energy ($D_{nn}$) |
|---|---|---|---|---|
| $Eu_2Ti_2O_7$ | -1.35K | 0.679 $\mu_B$ | -0.67K | +0.006K |

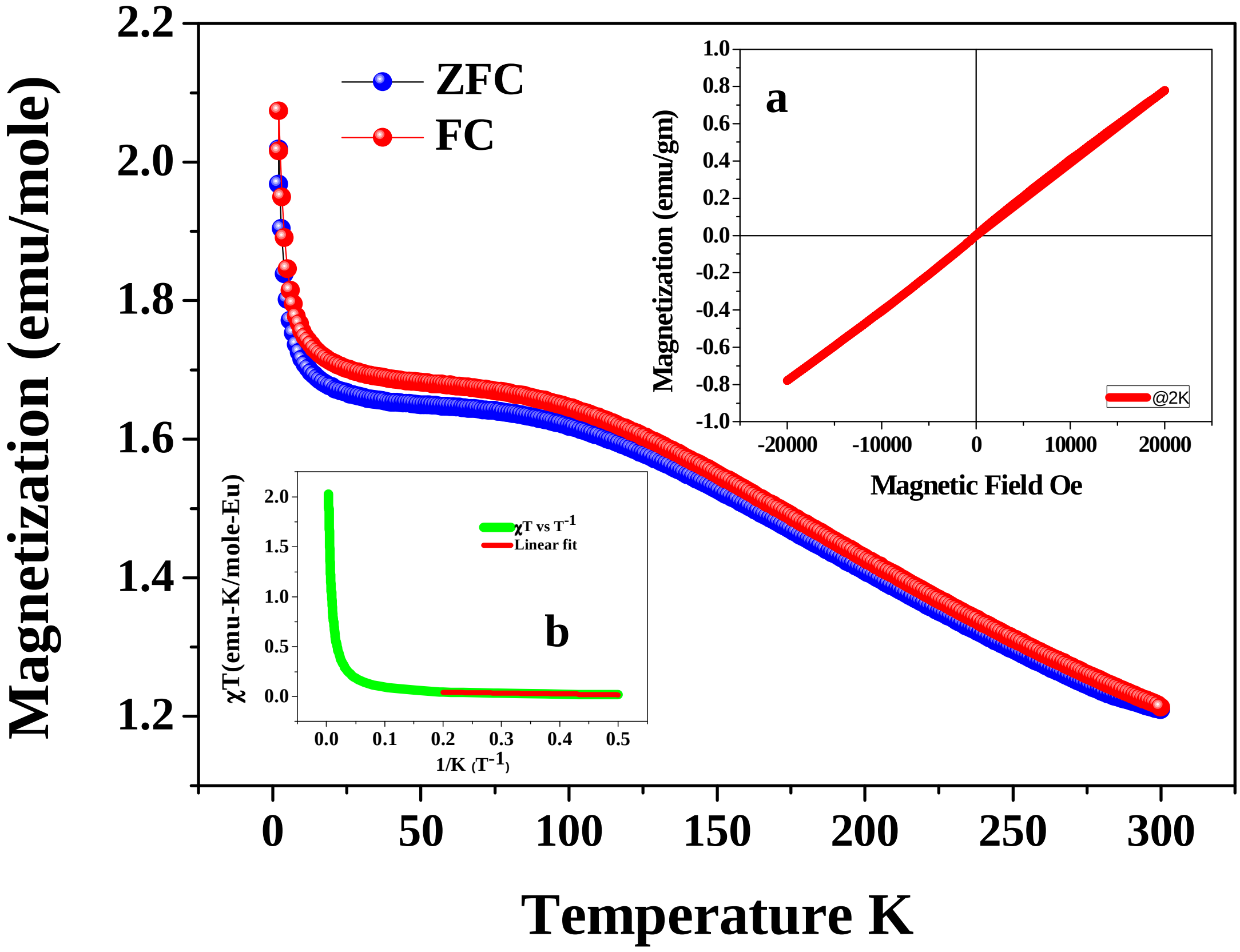

**Fig.2**

### *Ac magnetization study*

To study the spin relaxation process in $Eu_2Ti_2O_7$, ac-susceptibility was performed at the temperature range from 2K-80K at different frequencies. Surprisingly we observe a frequency dependent transition in the form of a dramatic drop in $\chi^{/}$ and the corresponding single peak as expected from Kramers-Kronig relations in $\chi^{//}$ while decreasing temperature below~ 35K [fig.3 (a &b)], although in the dc susceptibility no prominent transition was found. The freezing transition observed at such a high temperature is quite unusual and strange, since the structural disorder or the site randomness present in these pyrochlore materials is less than 1%. This directly creates contradiction to the well established spin-glass theory where disorder plays the main role in producing glassy behaviour [28]. However; the strong frequency dependence of these peak positions suggests the existence of a slow magnetic spin relaxation. The freezing temperature $\mathbf{T}_f$ is obtained from the clear sharp rise in $\chi^{//}$ which is correlated with the sharp drop in $\chi^{/}$ [11,19,29]. Below freezing temperature (i.e 35K), the dynamic of $Eu^{3+}$ spins response get slowed down as they cannot follow the time varying ac magnetic field, as a consequence, the drop in $\chi^{/}$ occurs [20,21]. Therefore, in the present case the relaxation time of the dynamic spins is longer than the measurement time (which is the inverse of the frequency) such that the system goes to "out of equilibrium" in that time scale, while the energy absorption by the relaxing spins is manifested by the rise of peak in the $\chi^{//}$. In this scenario, it is relevant to mention here that freezing transition was reported earlier in spin ice materials i.e. $Dy_2Ti_2O_7$ and $Ho_2Ti_2O_7$, $Ho_2Sn_2O_7$ etc. at very low temperature (~1K) [11,12,18,30]. For all these low temperature freezing transitions (~1K), the thermal energy barrier associated with the spin relaxation is~ 20K. But as mentioned in the introduction that only for the spin ice compound $Dy_2Ti_2O_7$; a further unexpected transition was observed at 16K which obeys Debye type exponential relaxation behaviour with thermal energy barrier 210K. This transition is unusual since it was not observed in any other spin ice compounds and thus is of particular scientific interest [11,20]. For the last two decades, intense research is going on to investigate the thermally activated transition at 16Kin the frustrated spin ice $Dy_2Ti_2O_7$. However, in our case, we have not observed any low temperature (<4K) transition but foundrelatively higher temperature (~34K, $\boldsymbol{f}$=700Hz) spin freezing making the situation much more interesting. To investigate the effect of a dc magnetic field on the spin freezing, we have studied the ac susceptibility with dc bias field of 10KOe. Figure 3(c & d) shows the graphs for $\chi^{/}$ and $\chi^{//}$as a function of temperature with dc bias H=10 KOe. It is clear from this figure that even after applying such high field, the freezing temperature $\boldsymbol{T_f}$ remains almost unchanged while the drop in $\chi^{/}$ decreases a bit and a little suppression occurs in corresponding $\chi^{//}$ peak.

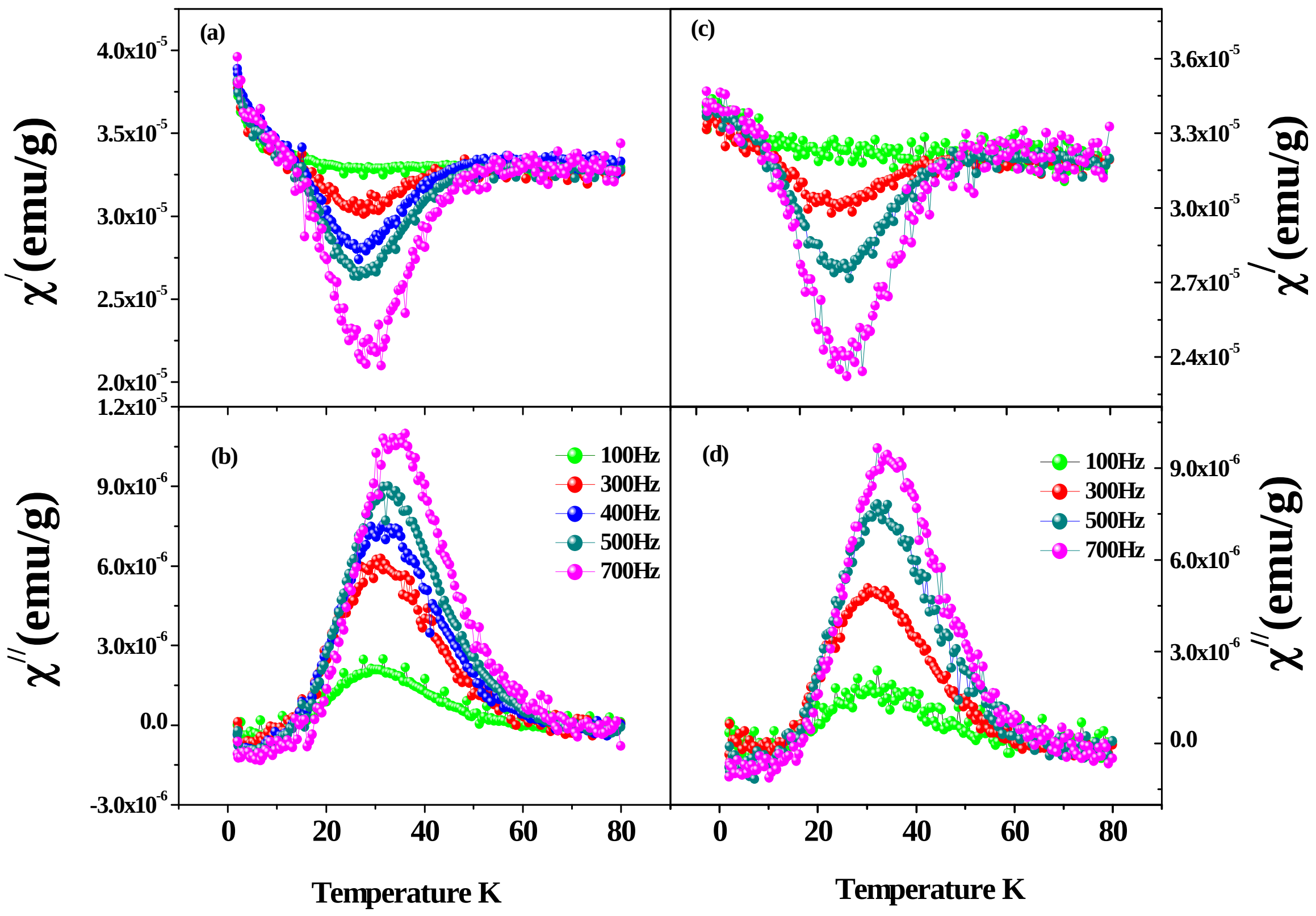

**Fig.3**

In fig. 4(a & b), we have shown the variation of $\chi^{/}$ and $\chi^{//}$ at different dc magnetic fields. Here it is very interesting to note that on application of further higher dc magnetic fields (2T and 3T), $\boldsymbol{T_f}$ shows clear shift towards higher temperature and the drop in $\chi^{/}$ gets suppressed (fig. 4(b)). Hence it is in strong contrast with typical spin glass transition. In typical spin glass, the freezing temperature gets shifted towards lower temperature with application of dc magnetic field as the applied field hinders the spins to freeze, but in this case, $\boldsymbol{T_f}$ shifts towards higher temperature which undoubtedly rules out the possibility of spin glass [31]. In spin ice $Dy_2Ti_2O_7$, though similar $\boldsymbol{T_f}$ shift was observed in its 16K transition but for this case $\chi^{/}$ itself got suppressed with dc field, hence qualitatively this transition seems different [12]. However, another way to check whether the present transition is of spin glass type, we have studied freezing temperature $\boldsymbol{T_f}$ as a function of frequency. For a typical spin-glass transition, it is characterized by a parameter $\boldsymbol{p} = \frac{\Delta T_f}{T_f \Delta(log f)}$ [12, 32,33] **,** where the value of $\boldsymbol{p}$ should be of the order of 0.01, where $\boldsymbol{T_f}$ is the freezing temperature at frequency $\boldsymbol{f}$. But in the present investigation, the value we obtained is~0.286 which is much greater than typical $\boldsymbol{p}$ value for spin glass. Therefore, this result confirms that the observed freezing transition is different from typical spin glass transition. It is though not very surprising because the observed transition is related to lattice geometry and not related to site-disorder which usually gives rise to spin glass behaviour. Again, to further characterize the observed transition, we have fitted the "frequency (*f*) dependence of freezing temperature ($T_f$)" by Arrhenius law $\boldsymbol{f = f_0 e^{-E_b/K_B T}}$ in the inset of fig.4(b); where $E_b$ is the thermal energy barrier for spin flipping and $f_0$ is a measure of the microscopic limiting frequency in the system, $K_B$ being the Boltzman constant [20,21]. The fit shows the thermal energy barrier ($E_b$) for pure $Eu_2Ti_2O_7$ is 339K and $f_0$ is of order MHz which is a reasonable value for spin flipping. The thermal energy barrier (339K), thus obtained is of the order of the crystal field (CF) energy level spacing between ground state $^7F_0$ and first excited state $^7F_1$ ( ~378K or 263 $cm^{-1}$) [17] which suggests the transition is thermally activated and the energy is of the order of single ion anisotropy energy for $Eu^{3+}$ ions [20,21].

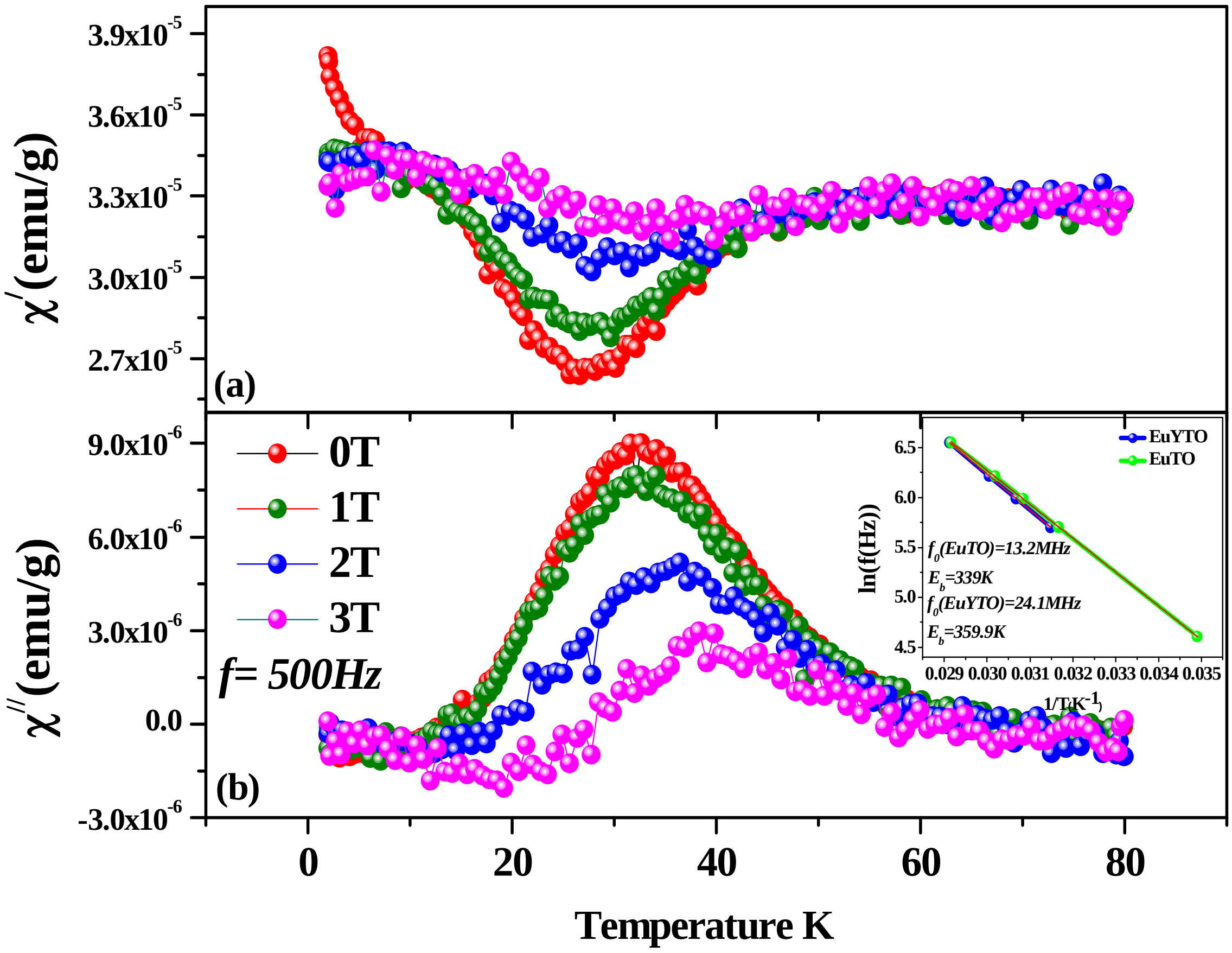

**Fig.4**.

In order to investigate the underlying freezing mechanism observed in pure $Eu_2Ti_2O_7$, the "frequency dependence of $\chi^{//}$ " (fig. 5) have been studied at different temperatures near and below $\boldsymbol{T_f}$ **(**<36K). The patterns of the "$\chi^{//}$ curves as a function of frequency ***f*"** change in a systematic manner with the temperature approaching towards the freezing temperature $\boldsymbol{T_f}$. The peaks of the $\chi^{//}$ curves get relatively sharper as it goes near the freezing temperature $\boldsymbol{T_f}$ , suggesting the underlying spin relaxation can presumably be attributed to the individual spin relaxation which is known as single ion freezing [12]. In general, a sharp peak indicates the distribution of spin relaxation times is very narrow around a "single characteristic time **τ"** which is defined as **τ=1/*f*** , where $\chi^{//}(f)$ having its maximum at ***f*** [12,20,29]. In contrast, in conventional spin-glass systems, the $\chi^{//}(f)$ shows a broad feature thus suggesting a spreading of relaxation times over several decades [12]. Thus the observed spin freezing is qualitatively different from spin-glass freezing. However, it is noteworthy to mention here that the sharpness of the $\chi^{//}(f)$ curves decreases a little bit as temperature cools below the freezing temperature $\boldsymbol{T_f}$ but still remains in a limit to show single spin relaxation process. Another way to characterize the observed spin freezing process is to fit the "$\chi^{//}$ (normalized to its maximum value) *vs.* the normalized frequency $\frac{f}{f_{peak}}$ " curves (top inset of fig. 5) by ***Casimir-du Pŕe*** relation which is typically used as to predict a single ion relaxation process [12,20]. The relation is $\boldsymbol{\chi''(f) = f\tau[(\chi_T - \chi_S)/(1 + f^2\tau^2)]}$, where $\boldsymbol{\chi_T}$ is isothermal susceptibility in the limit of low frequency and $\boldsymbol{\chi_S}$ is the adiabatic susceptibility in the limit of high frequency. It is observed that our experimentally obtained curves below and near $\boldsymbol{T_f}$ (27.16K and 33.16K), are relatively narrower than the theoretically fitted data, suggesting the observed spin freezing is caused by individual spin relaxation process. Further, for a single spin relaxation mode, it is theoretically predicted that Cole-Cole (Argand) plot of "$\chi^{//}$ as a function of $\chi^{/}$", should show a semicircular nature [12]. In (bottom) inset of fig.5, the Cole- Cole plots have been shown at different temperatures (below and near $\boldsymbol{T_f}$). Interestingly, the curves clearly trace the semicircular path thus indicating the presence of the single spin relaxation process in the observed spin freezing. The change in the peak positions of $\chi^{//}$, implies slight change in the relaxation times as temperature decreases.

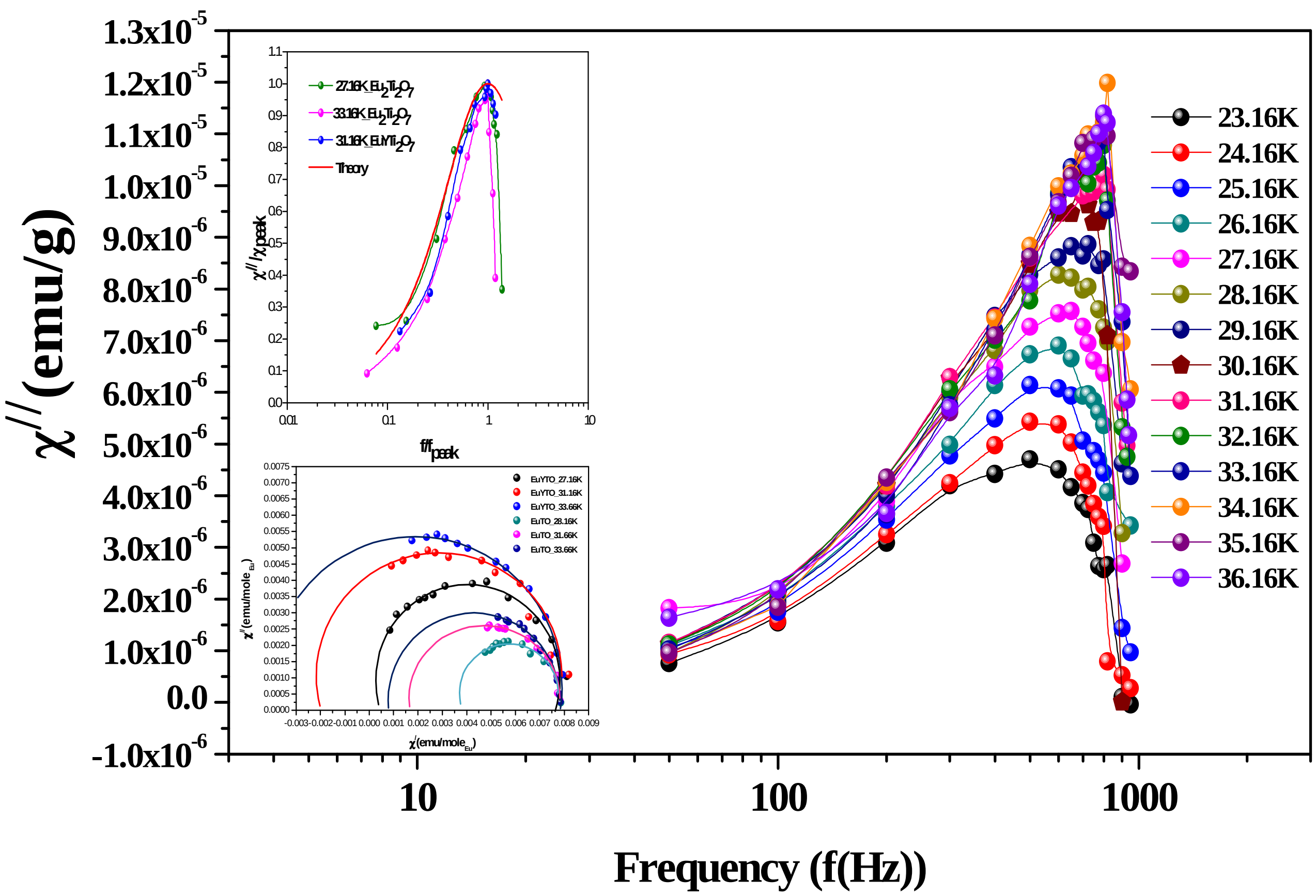

**Fig. 5:**

In the present scenario, to confirm the nature of this spin freezing transition, we investigated the effect of non-magnetic dilution in our system. We have replaced 50% of $Eu^{3+}$ ions by non-magnetic $Y^{3+}$ ions to change the local environment of Eu ions such that in each of the corner shared tetrahedra there are two Eu ions and two Y ions effectively. Hence, this non-magnetic dilution causes increase in effective distance between Eu ions thus lowering the local Eu-Eu spin correlations. It is obvious that if the observed spin freezing (T<35K) requires spin-spin correlation, then the non-magnetic dilution should suppress the spin freezing. Interestingly, in the ac susceptibility (fig.6) of non-magnetic diluted sample $EuYTi_2O_7$ (EuYTO), the spin freezing transition is observed to become more pronounced instead of being diminished. This observation confirms the single ion nature of the observed spin freezing, where local spin-spin correlation is not required at all [21]. The possible explanation for the enhancement of the spin freezing with non-magnetic dilution can be given on the basis of the change in the local environment of the participating $Eu^{3+}$ spins. In fig. 6(c), the increased drop in $\chi^{/}$ indicates a further increase in the characteristic relaxation time of spin flipping with $Y^{3+}$ substitution i.e. the spins take longer time to relax. From, top inset of fig. 5, it is clear that even after 50% dilution, the $\chi^{//}(\boldsymbol{f})$ remains narrow enough to support single ion freezing. The calculation of thermal energy barrier ($E_b$) by Arrhenius plot for EuYTO gives a value~360K which is greater than that of EuTO (339K) [inset of fig. 4(b)]. The increased $E_b$ causes more delay in the spin relaxation, thus enhancing the spin freezing with dilution. The small increase in thermal energy barrier $E_b$ is seemingly caused by slight alteration of lattice constant and electronic structures of the system [21]. Bottom inset of fig. 5, shows even after 50% non-magnetic dilution, the Cole-Cole plot follows closely semicircular path, suggesting single ion process. Therefore, the observed transition is fundamentally a single ion process and is not affected by spin-spin correlation. However, a noticeable difference between spin freezing (16K) observed in $Dy_2Ti_2O_7$ (DTO) and the spin freezing observed in EuTO is that for DTO freezing temperature ($\boldsymbol{T_f}$) increases with dilution but for EuTO, upto 50% dilution, it remains almost unchanged (fig. 6(d)). However, it deserves further study to fully understand the underlying physics.

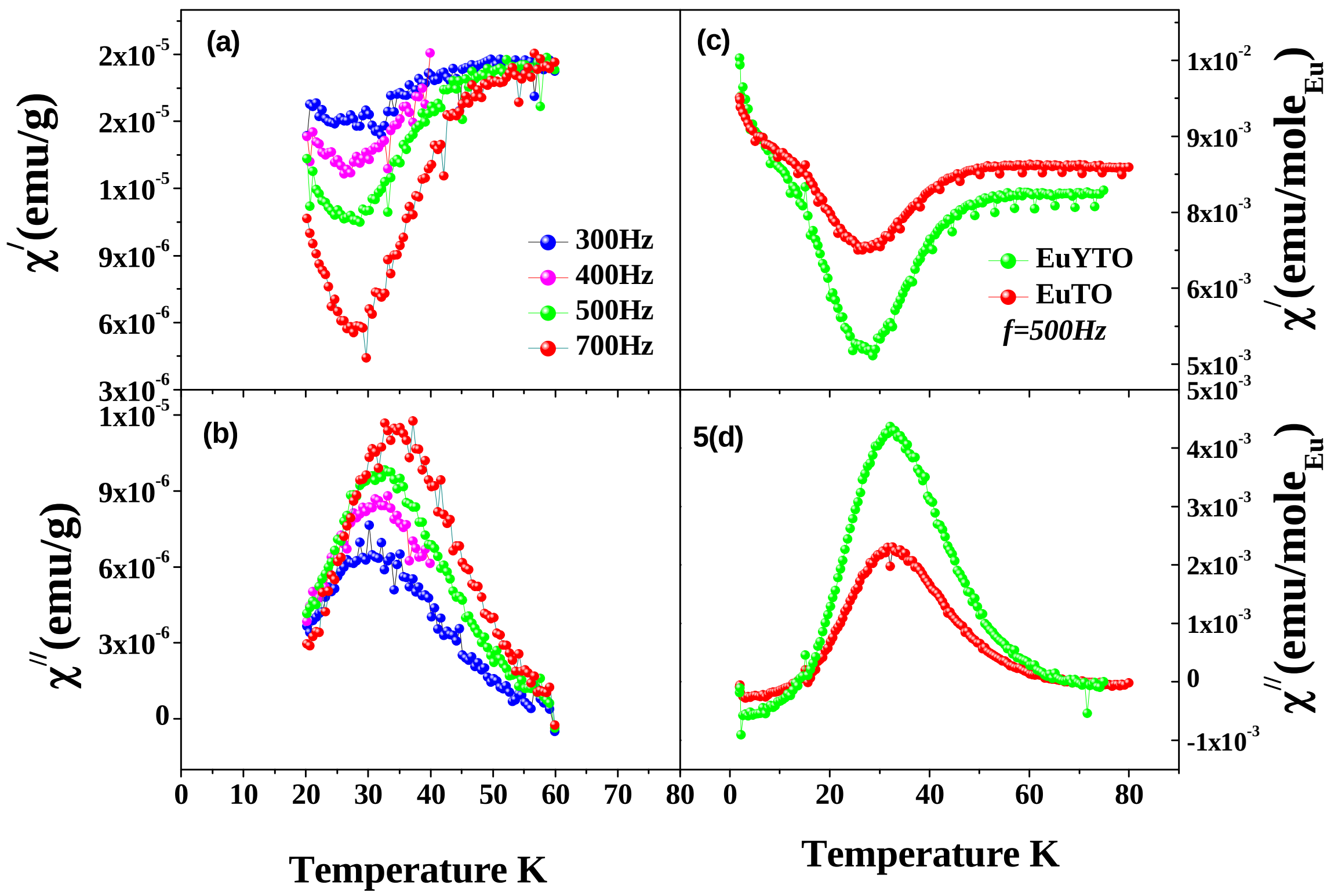

**Fig. 6:**

## Conclussion

In summary, we have performed a systematic study of dc and ac magnetization of pyrochlore $Eu_2Ti_2O_7$. The new observation of prominent spin freezing at such a higher temperature (< 35K) is quite surprising in this site ordered pyrochlore. Detailed analysis of the observed freezing transition have been carried out in many different ways to investigate the spin dynamics behind this freezing. The thermal energy barrier ($E_b$) as extracted from the frequency dependence of ac $\chi^{//}$ peak positions, is found to be of the order of single ion anisotropy energy i.e the energy level spacing of CF levels. By nature, the observed transition differs from common spin glass behaviour in many ways, though this is not very surprising since the pure pyrochlore materials offer very little structural disorder(<1%). In common spin glass materials, the presence of large scale structural and chemical disorders offer the spin frustration, where the spins are almost isotropic in nature. Whereas for pyrochlore $Eu_2Ti_2O_7$, the observed freezing is caused by local geometrical spin frustration where the spins have anisotropy (SIA) which offers them less freedom for movement. Hence, this spin freezing transition ( T<35K) is fundamentally different by its origin from other glassy transitions. Analysis by Cole-Cole plot and Casimir-du Pŕe relation, further confirmed the single ion relaxation process to be involved in the observed spin freezing. Furthermore; despite the dilution (50%) of magnetic $Eu^{3+}$ ions by purely non-magnetic $Y^{3+}$ ions; the spin freezing became more pronounced instead of being suppressed. Thus such non-magnetic dilution rules out the probability of need for "local spin-spin correlation" in this spin freezing process; thereby suggesting a "single spin relaxation mechanism" in this freezing process. More interestingly, here the temperature at which the spin freezing is observed (<35K), is quite higher than the temperatures at which spin freezing transitions were reported earlier in $Dy_2Ti_2O_7$ (<4K and 15K). Thus, the observed spin freezing transition is significantly unusual by its nature and further investigations may explore the understanding of spin dynamics in such site-ordered geometrically frustrated systems. Other theoretical models or neutron experiments may help to confirm its origin further.

## ACKNOWLEDGEMENTS

We are thankful to the Central Instrumentation Facility Centre, Indian Institute of Technology (BHU) for their assistance in magnetic measurements.

## REFERNECES:

Figure Captions:

**Fig.1**: X-Ray diffraction pattern with Rietveld refinement for $Eu_2Ti_2O_7$. The inset showing the structure of pyrochlore $Eu_2Ti_2O_7$. Here the blue, red and green spheres representing the position of $Eu^{3+}$, $Ti^{4+}$ and $O^{2-}$ ions respectively.

**Fig. 2**:The temperature variation of magnetization (ZFC and FC) curves at H=100Oe. Inset a showing isothermal magnetization as a function of fields at T=2K. Inset b showing "**$\chi$T vs** $\boldsymbol{\frac{1}{T}}$" curve and its linear fit for the range T=2-5K.

**Fig. 3:** : Temperature variation of ac susceptibility $\chi^{/}$ and $\chi^{//}$ at dc fields H=0 Oe(fig. a & b)and 10KOe for $Eu_2Ti_2O_7$ for different frequencies (fig. c & d).

**Fig. 4:** Temperature dependence of $\chi^{/}$ and $\chi^{//}$ at different dc fields at 500Hz. The inset showing Arrhenius plot for $Eu_2Ti_2O_7$ (EuTO) and $EuYTi_2O_7$ (EuYTO).

**Fig. 5:** : Frequency dependence of $\chi^{//}$ at different temperatures for pure $Eu_2Ti_2O_7$. Inset(top) showing the normalized $\chi^{//}$ as a function of $\boldsymbol{f/f_{peak}}$ and its theoretical fitting by Casimir du pŕe relations at different temperatures for pure and nonmagnetic diluted samples. Inset (bottom) showing Cole Cole (Argand) plot of "$\chi^{//}$ Vs $\chi^{/}$" at different temperatures for pure and diluted samples.

**Fig. 6:** : (a & b) showing temperature variation of ac susceptibility $\chi^{/}$ and $\chi^{//}$ for diluted sample EuYTO. Fig.(c and d) show comparison of temperature dependence of $\chi^{/}$ and $\chi^{//}$ for pure and diluted samples at 500Hz.